\documentclass[aps,prb,amsmath,floatfix,twocolumn,showpacs]{revtex4}

\usepackage{amsmath}

\begin{document}

\title{Generalized Galitskii approach for the vertex
function of a Fermi gas with resonant interaction}

\author{A. Vagov}
\author{H. Schomerus}
\affiliation{Department of Physics, Lancaster University,
Bailrigg, Lancaster, LA1 4YB, UK}
\author{A. Shanenko}
\affiliation{Departement Fysica, Universiteit Antwerpen,
Groenenborgerlaan 171, B-2020 Antwerpen, Belgium}

\date{\today}

\begin{abstract}
We present a generalized Galitskii approach  for the
Bethe-Salpeter equation for the two-particle vertex function of a
Fermi system with resonant interaction by accounting for the
resonant state in the scattering potential and utilizing the
universal form of the resonant scattering amplitude. The procedure
can be carried out both for the normal as well as for the
condensate state. In both cases, the vertex function in the
vicinity of the resonance is shown to formally coincide with that
obtained for a weakly attractive Fermi gas. Thus we justify the
popular calculational framework in which results for the weakly
attractive Fermi gas are formally extrapolated into the domain of
strong coupling, and further to the repulsive side of the
resonance, where molecular states are formed.
\end{abstract}

\maketitle

\section{Introduction}

The problem of an interacting Fermi system with strong resonant
coupling is relevant to many areas of physics, in particular,
ultracold gases \cite{Review1,Review2,Review3} and high $T_c$
superconductors. \cite{Review1} Experimental insight into this
problem has recently been obtained  using ultracold atomic $^6$Li
and $^{40}$K gases, \cite{Review2,Review3} in which the $s$-wave
scattering length $a$ can be tuned from positive (repulsive) to
negative (attractive) values by crossing a Feshbach resonance, at
which the scattering length diverges. These experiments started
with the successful formation of long-living paired fermions  in
the strong-coupling regime, \cite{Molecules} which were
subsequently seen to condense into a Bose-Einstein condensate
(BEC) of bound molecular states. \cite{BEC1,BEC2,BEC3} Soon
afterward the interaction was tuned to the attractive side,
\cite{BECBCS1,BECBCS2,BECBCS3,BECBCS4} and it was established that
the condensate crosses over to a BCS superfluid of extended Cooper
pairs. \cite{BCS1,BCS2,BCS3,BCS4,BCS5,BCS6}

The first theoretical studies of the BCS-BEC crossover were
initiated not long after the development of the BCS theory of
superconductivity itself, when it was noticed that the nonzero
solution of the BCS gap equation, complemented with the condition
of a constant number of particles and regularized to eliminate an
ultraviolet divergency, can be smoothly extrapolated to the domain
of positive scattering lengths (without diverging at the
resonance), where it describes molecular states.
\cite{meanfield1,meanfield2,meanfield3,meanfield4}

This mean-field interpolation between the two types of condensates
is believed to be most accurate at $T=0$, where bosonic
excitations of paired fermions (Cooper pairs or molecular states
of finite momentum) are not populated. Subsequent works
established that the pair excitations  induce fluctuation
corrections which modify the thermodynamics especially in the
strong-coupling limit.
\cite{beyondmeanfield00,beyondmeanfield2,beyondmeanfield3,Ohashi2002,beyondmeanfield5,CombescotPRA73}
In many of these works the fluctuation corrections are treated
within the self-consistent $T$-matrix approximation which involves
the ladder approximation for the vertex function.

The calculations beyond the mean field approximation in the BCS
theory were initiated by Gor'kov and Melik-Barkhudarov;
\cite{Gorkov} similar types of corrections were also applied to
the BCS-BEC crossover. \cite{Heiselberg1} Other recent approaches
combine the ladder approximation with the fully self-consistent
solution for the single particle Green function,
\cite{beyondmeanfield4} and account for the scattering of the
weakly bound composite bosons. It should be noted that the
scattering length for the two molecules {\em in vacuo} can be
calculated exactly, \cite{Shlyapnikov1} providing a useful
guidance for the many-body calculations. Furthermore, the BEC-BCS
crossover can also be studied numerically in Monte Carlo
simulations, \cite{MonteCarlo1,MonteCarlo2,MonteCarlo3} or via the
renormalization group analysis with $\varepsilon$ and $1/N$
expansions at the unitary point.
\cite{Renormgroup2,Renormgroup3,Renormgroup4,Renormgroup1,Renormgroup5}

Comparison with experiment and Monte-Carlo calculations reveal
that the ladder approximation captures the qualitative features
for various physical quantities in the BCS-BEC crossover far
better than it could be
expected.\cite{beyondmeanfield8,beyondmeanfield6} This is also
astonishing since the corresponding Bethe-Salpeter equation is
conventionally derived by assuming the constant approximation for
the scattering matrix,\cite{FetterWalecka} which does not
explicitly account for the resonant state. This approximation is
strictly valid only for a weakly attracting Fermi gas. In order to
study the strong-coupling limit, the resulting expressions are
formally extrapolated, and these expressions are also used on the
repulsive side of the resonance. Curiously, they then capture
molecular states of the correct energy, and also recover the
scattering length for the molecules in the Born approximation.
Moreover, the results turn out to be nondivergent at resonance.
These observations have motivated several works which address the
justification of solving the Bethe-Salpeter equation for the
strong coupling case and in the presence of bound molecular
states. In the context of high-$T_c$ two-dimensional
superconductors the validity of the extrapolation procedure for
the strong coupling regime was discussed by Randeria {\em et al.}
\cite{Randeria}, while for the ultracold Fermi gases the
properties of the resonant scattering amplitude were used by
Combescot \cite{CombescotPRL2003} to search for the universal form
of the scattering matrix which would be applicable close to the
resonance while recovering both limiting cases far from it.

In this work we solve the Bethe-Salpeter equation for the
two-particle vertex function of a Fermi system near the resonance,
and show that the resulting expression coincides with the result
of the extrapolation and regularization procedure. Our derivation
proceeds via an extension of the Galitskii
formalism,\cite{FetterWalecka} which reformulates the
Bethe-Salpeter equation by using the scattering amplitude instead
of the scattering potential. In its original version, the
Galitskii formalism is unsuitable to describe resonant scattering,
since it does not explicitly account for the resonant state and as
a result treats the case of positive scattering length $a$ as a
repulsive Fermi system. In our derivation we do not use any
assumptions beyond the condition that the resonant state induces a
broad Feshbach resonance. The state is eliminated by using the
completeness and orthogonality of the set of all scattering
states. This procedure can be carried out both for the normal
state as well as for the condensate state, and leads to a fully
analytical solution of the Bethe-Salpeter equation which is
inherently valid in the vicinity of the resonance, as well as in
the domain of the positive scattering length. The resulting
expressions of this formal derivation turn out to be identical to
the BCS extrapolation scheme with the renormalized contact
interaction. Furthermore, our derivation constitutes a
reorganization of the ladder approximation which demonstrates that
the relevant expansion parameter does not diverge at resonance.
Thus we establish a firm basis for the common conceptual framework
which is used both in the mean field description of the BCS-BEC
crossover, as well as in many calculations accounting for the
fluctuation corrections.

This paper is organized as follows. Section \ref{sec:model} defines
the microscopic Fermi model on which all considerations are based.
In Section \ref{sec:normal} we present the modified Galitskii
formalism for the Fermi gas in the normal state. In Section
\ref{sec:condensate} we present the derivation for the Fermi system
in the condensate state, and also briefly discuss the gap equation.
The results of this paper are summarized in Section
\ref{sec:conclusions}.

\section{\label{sec:model}Single-channel model for the resonantly
interacting Fermi gas}

In order to describe an ultracold gas of fermionic atoms in the
vicinity of a Feshbach broad resonance we use the standard
single-channel Hamiltonian
\begin{align}
H &= \sum_{\alpha}\int dx \left\{ \psi_\alpha^\dag({\bf r})\left(
-\frac{\hbar^2\nabla^2 }{2m}-\mu \right)  \psi_\alpha({\bf
r})\right\}\notag \\
&+ \frac{1}{2}\sum_{\alpha,\beta} \int d{\bf r} d{\bf r}^\prime
\psi_\alpha^\dag({\bf r})\psi_\beta^\dag({\bf
r}^\prime)U_{\alpha\beta}({\bf r}-{\bf r}^\prime) \psi_\beta({\bf
r}^\prime) \psi_\alpha({\bf r}), \label{0}
\end{align}
where $\psi_\alpha({\bf r})$ is a fermionic field operator for a
particle with spin $\alpha$. In the following we work in the units
$m\equiv 1$, $\hbar\equiv 1$, and also set Boltzmann's constant
$k_B\equiv 1$. The effective scattering potential
$U_{\alpha\beta}(\bf r)$ describes resonant scattering of
particles of opposite spin in the $s$-wave channel. In the
vicinity of a broad resonance with effective interaction radius
$r_0\ll k_F^{-1}$ (where $k_F$ is the Fermi momentum), the
scattering amplitude in the $s$-wave channel takes the universal
form \cite{Landau}
\begin{align}
\label{1-7a} f({\bf k}) \approx -\frac{1}{\eta +i |{\bf k}|},
\end{align}
where $\eta=a^{-1}$ is the inverse scattering length. The
scattering length $a$ diverges at resonance, $\eta =0$.  For
$\eta<0$ the resonant level lies in the continuum, and the
interaction is attractive.  For $\eta>0$ the resonant level turns
into a bound state, and the interaction is repulsive.  In the BCS
limit $\eta \rightarrow -\infty$, the interaction mediates the
formation of Cooper pairs, which are weakly bound in momentum
space.  In the BEC limit $\eta \rightarrow +\infty$, the
interaction mediates the formation of molecular states, which are
weakly bound in real space. Because these notions refer to the
condensate phase, we more broadly speak of the Fermi limit for
$\eta\to-\infty$, and of the Bose limit for $\eta\to+\infty$.

It should be stressed that the Hamiltonian (\ref{0}) is purely
fermionic, and can be contrasted to the more detailed Fermi-Bose
models which explicitly account for a bound molecular state in the
closed channel.
\cite{Holland2001,Ohashi2002,Milstein2002,Falco2004} A great deal
of work has been carried out in the past to give a detailed
justification of the validity of both models and, in particular,
to show their equivalence in the strong-coupling regime of a broad
Feshbach resonance.
\cite{Kokkelmans2002,Bruun2004,Diener2004,Simonucci2005,Falco2007,Stecher2007}
The universality arises for a combination of two facts. Firstly,
at $\eta=0$  the scattering length diverges, which renders the
Fermi energy $E_F=k_F^2/2$  as the only relevant energy scale for
the thermodynamics. \cite{Ho2004} Secondly, in this unitary limit
the contribution of the closed channel becomes negligible, which
has been directly tested in experiments on $^6$Li,
\cite{ModelExp2} even though the situation in $^{40}$K may be a
matter of debate. \cite{Parish2005,Szymanska2005}

\section{\label{sec:normal}Modified Galitskii formalism for
resonant scattering} We first assume that the Fermi gas is in the
normal state.  In the ladder approximation, the vertex function
$\Gamma(p_1,p_2;p_3,p_4)$ is then determined by the Bethe-Salpeter
equation
\begin{align}
&\Gamma(p_1,p_2;p_3,p_4) = u({\bf p}_1 - {\bf p}_3) - T\sum_{q_{0}}
\int \frac{d {\bf q}}{(2\pi)^{3}} u({\bf
q}) \notag \\
&\times  G(p_1-q) G(p_2 + q)   \Gamma(p_1-q, p_2+q; p_3, p_4),
\label{1-1}
\end{align}
where spin indices are omitted for clarity,  $p_i=(p_{i,0},{\bf p})$
is the 4D energy-momentum vector of an incoming ($i=1,2$) or
outgoing ($i=3,4$) particle, $u({\bf q})$ is the momentum
representation of the interaction potential, and
\begin{align}
\label{1-2} G(q)=\frac{1}{q_0 -\xi_{{\bf q}}}, ~~~\xi_{\bf q} =
\frac{{\bf q}^2}{2} - \mu,
\end{align}
is the Green function of a noninteracting Fermi system. We use the
notation $q=(q_0, {\bf q})$, where  ${\bf q}$ denotes the
three-dimensional momentum vector and $q_0= i \pi T (2n+1)$ is the
Matsubara frequency.

In solving Eq.\ (\ref{1-1}) one follows the conventional
strategy\cite{FetterWalecka}, developed for the $T=0$ case, where
the potential $u({\bf q})$ is eliminated in favor of the vacuo
vertex function $\Gamma_0$, defined as the solution of Eq.
(\ref{1-1}) for two particles in absence of all the other particles,
i.e., by setting $\mu=0$. An extension of this strategy for the
finite temperature case is conveniently achieved by a subsequent
analytic continuation of the energy argument in $\Gamma_0$ as
\begin{equation}
E = g_0 + 2 \mu - {\bf g}^2/4. \label{eq:edefinition}
\end{equation}
Hereafter, we denote the total momentum of the scattered particles
as $g=p_1+p_2=(g_0,{\bf g})$, while $p=(p_1-p_2)/2$ and
$p^\prime=(p_3-p_4)/2$ are the relative momenta before and after the
scattering event, respectively.

Eliminating $u({\bf p})$ in Eq. (\ref{1-1}), the Bethe-Salpeter
equation takes the form
\begin{align}
\Gamma({\bf p},{\bf p}^\prime, g) &= \Gamma_0({\bf p},{\bf
p}^\prime, g)  \notag \\ &-
 \int \frac{d^3 {\bf q}}{(2 \pi)^3}
\Gamma_0({\bf p},{\bf q}, g) K({\bf q},g) \Gamma({\bf q}, {\bf
p}^\prime, g). \label{1-3}
\end{align}
The kernel $K$ is defined as
\begin{align}
K( {\bf q}, g)  &= T \sum_{q_0} G\left(\frac{{\bf g}}{2} - {\bf q},
g_0-q_0\right) G\left(\frac{{\bf g}}{2}+q,q_0 \right) \notag \\
&+ \frac{1}{E - {\bf q}^2}. \label{1-4}
\end{align}
Using Eq. (\ref{1-2}) one obtains an explicit expression of the
kernel,
\begin{align}
 &K({\bf q},g) =  \frac{n_{F}(\xi_{+}) +n_{F}(\xi_{-})}{E - {\bf q}^2},
\label{1-5}
\end{align}
where $n_{F}(\varepsilon) = [\exp( \varepsilon/ T ) +1]^{-1}$ is the
Fermi function and
\begin{equation}
\xi_\pm =\frac{1}{2}\left(\frac{{\bf g}}{2} \pm {\bf
q}\right)^{2}-\mu. \label{eq:xidefinition}
\end{equation}

In the Galitskii approach (see, e.g., Ref.
\onlinecite{FetterWalecka}), the vacuo vertex function $\Gamma_{0}$
is found by relating it to the two-momentum scattering amplitude
$f({\bf p},{\bf k})$. However, the original formalism does not
account for a possible bound molecular state in the scattering
potential. In the following we assume that the scattering potential
$u({\bf q})$ allows for a single bound state, with eigenfunction
$\phi(x)$ and eigenvalue $\lambda$.

In the first step,  $\Gamma_0$ is written as an integral
\begin{align}
\Gamma_0 ({\bf p},{\bf p}^\prime, g) = \int \frac{d^3 {\bf
q}}{(2\pi)^3} u({\bf q}) \chi ({\bf p}-{\bf q},{\bf p}^\prime,g),
\label{A-1}
\end{align}
where the scattering state $\chi ({\bf p},{\bf p}^\prime,g)$
satisfies the equation
\begin{align}
&(E - {\bf p}^2 +i 0) \chi ({\bf p},{\bf p}^\prime,g)- \int
\frac{d^3 {\bf q}}{(2
\pi)^3} u({\bf q})\chi ({\bf p}-{\bf q},{\bf p}^\prime,g)\notag \\
&=(2 \pi)^3(E - {\bf p}^2 +i 0)\delta({\bf p}-{\bf p}^\prime).
\label{A-2}
\end{align}
The left-hand side of this equation is equivalent to a
Schr\"odinger equation in momentum space. Therefore, $\chi ({\bf
p},{\bf p}^\prime,g)$ can be written using the complete set of
solutions of the Schr\"odinger equation,
\begin{align}
&\chi ({\bf p},{\bf p}^\prime,g)=(E - {\bf p}^{\prime 2} +i 0) \int
\frac{d^3 {\bf k}}{(2 \pi)^3} \frac{\psi_{\bf k}({\bf p})
\psi_{\bf k}^*({\bf p}^\prime)}{E - {\bf k}^2 +i 0} \notag \\
&+ \frac{E - {\bf p}^{\prime 2} +i 0 }{E -2\lambda +i 0}\phi({\bf
p}) \phi^*({\bf p}^\prime), \label{A-3}
\end{align}
where $\psi_{\bf k}({\bf p})$ is the Fourier transform of the
scattering state $\psi_{\bf k}({\bf r})$ with momentum ${\bf k}$.
The last term of this expression explicitly accounts for the bound
state $\phi({\bf p})$. The states $\psi_{\bf k}({\bf p})$ can be
expressed via the two-momentum scattering amplitude $f({\bf
p},{\bf k})$,
\begin{align}
\psi_{\bf k}({\bf p}) = (2 \pi)^3 \delta({\bf p}-{\bf
k})+\frac{f({\bf p},{\bf k})}{ {\bf k}^2-{\bf p}^2 +i 0}.
\label{A-4}
\end{align}
Inserting this expression into Eq. (\ref{A-3}) one obtains
\begin{align}
\chi({\bf p},{\bf p}^\prime,g) & = \psi_{{\bf p}^\prime}({\bf p})+
\int \frac{d^3 {\bf k}}{(2 \pi)^3} \psi_{{\bf k}}({\bf p}) f^*({\bf
p}^\prime,{\bf k}) \notag \\
&\times \left( \frac{1}{E - {\bf k}^2 +i 0}+ \frac{1}{{\bf
k}^2 - {\bf p}^{\prime 2} -i 0} \right) \notag \\
&+ \frac{E- {\bf p}^{\prime 2} +i 0}{E -2\lambda +i 0}\phi({\bf p})
\phi^*({\bf p}^\prime) . \label{A-5}
\end{align}
After substituting Eq.  (\ref{A-5}) into Eq.  (\ref{A-1}) one
obtains $\Gamma_{0}$ as a sum of two terms
\begin{align}
    \Gamma_{0} \left( {\bf p}, {\bf p}^{\prime}, g \right) =
    \Gamma_{0}^{G}\left( {\bf p}, {\bf p}^{\prime}, g \right) +
    \Gamma_{0}^{b}\left( {\bf p}, {\bf p}^{\prime}, g \right).
    \label{1-6}
\end{align}
The first term in Eq.\ (\ref{1-6}) recovers the original Galitskii
result
\begin{align}
    \Gamma_{0}^{G} \left( {\bf p}, {\bf p}^{\prime}, g \right)
    &= f \left(
{\bf p}, {\bf p}^{\prime} \right) + \int \frac{d^3 {\bf
k}}{(2\pi)^{3}} f \left( {\bf p}, {\bf k} \right) f^{*} \left({\bf
p}^{\prime}, {\bf k} \right) \notag \\
& \times \left\{ \frac{1}{E-{\bf k}^{2}+i 0} + \frac{1}{{\bf
k}^{2}-{\bf p}^{\prime 2}-i 0} \right\}.
\label{1-6a}
\end{align}
The second term in Eq.\ (\ref{1-6}) originates from  the bound
state and is given by
\begin{align}
\Gamma_{0}^{b} \left( {\bf p}, {\bf p}^{\prime}, g \right) = \frac{E
-{\bf p}^{\prime 2}}{E -2\lambda} \int \frac{d^{3} {\bf
q}}{(2\pi)^{3}} u({\bf q}) \phi( {\bf p} - {\bf q}) \phi^* ({\bf
p}^{\prime}). \label{A-6}
\end{align}
This expression still explicitly depends on the bound state, and
at first glance it appears that detailed knowledge of $\phi ({\bf
p})$ is necessary in subsequent calculations. However, $\phi ({\bf
p})$ can be eliminated by relating it to $f({\bf k},{\bf p})$. To
establish this relation we utilize the Schr\"odinger equation for
$\phi ({\bf p})$ in the momentum representation
\begin{align}
(2\lambda - {\bf p}^2) \phi({\bf p}) = \int \frac{d^{3} {\bf
q}}{(2\pi)^{3}} u({\bf q}) \phi( {\bf p}-{\bf q}) \label{A-6a}
\end{align}
and the completeness relation
\begin{align}
\int \frac{d^3 {\bf k}}{(2 \pi)^3} \psi_{\bf k}({\bf p})\psi_{\bf
k}^*({\bf p}^\prime) + \phi({\bf p})\phi^*({\bf p}^\prime) =(2
\pi)^3 \delta ({\bf p} -{\bf p}^\prime). \label{A-8}
\end{align}

Multiplying both side of Eq. (\ref{A-6}) by $\phi ({\bf p})$, then
using Eq.\ (\ref{A-8}) and the definition (\ref{A-4}) one obtains
the correction of the Galitskii result due to the contribution of
the molecular state as
\begin{align}
\Gamma_{0}^{b} \left( {\bf p}, {\bf p}^{\prime}, g \right)
    &= \frac{E-{\bf p}^{\prime 2}}{E-2\lambda}\frac{2\lambda - {\bf
p}^{2} }{{\bf p}^{2} - {\bf p}^{\prime 2} - i 0} \bigg\{ \int
\frac{d^3 {\bf k}}{(2\pi)^{3}} f \left( {\bf p}, {\bf k}
\right)  \notag \\
&\times f^{*} \left({\bf p}^{\prime}, {\bf k} \right) \left(
\frac{1}{{\bf k}^{2}-{\bf p}^{\prime 2}-i 0}-\frac{1}{{\bf k}^2-{\bf
p}^{2}+i 0}
\right) \notag \\
&  +f({\bf p},{\bf p}^\prime) - f^*({\bf p}^\prime , {\bf p})
\bigg\}.
\label{1-6b}
\end{align}

Equations (\ref{1-6}), (\ref{1-6a}) and (\ref{1-6b}) reduce the
calculation of the vacuo vertex to that for the two-momentum
scattering amplitude $f({\bf p},{\bf k})$, defined by Eq.
(\ref{A-4}). This amplitude is obtained by solving the
Lippmann-Schwinger equation.  In the weak-scattering limit,
$f({\bf p},{\bf k})$ can be approximated by a constant $4
\pi/\eta$. Following Eq.\ (\ref{1-6}), this leads to a constant
approximation for $\Gamma_0$, as well. Since the vertex function
then does not possess any poles, the constant approximation cannot
be used to describe the bound states. In a more accurate analysis,
applicable close to the resonance as well as in the limits
$\eta\rightarrow \pm \infty$, one can utilize the fact that in the
vicinity of a resonance the two-particle scattering amplitude for
$s$-wave scattering $f({\bf p},{\bf k})$ assumes as similar
universal form as the single-particle scattering amplitude $f({\bf
k})$, \cite{Landau}
\begin{align}
\label{1-7b} f({\bf p},{\bf k}) =\frac{4\pi}{\eta +i |{\bf k}|},
\end{align}
where in general $\eta$ depends on ${\bf p}$ and ${\bf k}$. For
small momenta, $\eta$ approaches a constant equal to the inverse
scattering length. The leading corrections are quadratic,
$\propto {\bf k}^{2}$ and $\propto
 {\bf p}^{2}$, but the coefficients for these terms are
proportional to the effective radius of the scattering potential
$r_{0}$. \cite{Landau,FetterWalecka}  Under the condition of a broad
resonance one has $r_{0}k_{F}\ll 1$. Therefore, these terms can be
neglected and $\eta$ in Eq. (\ref{1-7b}) can be assumed constant,
equal to the inverse scattering length. On the molecular side of the
resonance, where the potential permits a bound state, the inverse
scattering length is furthermore related to the bound state energy
via $2\lambda=-\eta^2$.\cite{Landau}

It should be noted that Eq. (\ref{1-7b}) reduces to the correct
constant expressions in the weak-coupling limits $\eta \rightarrow
\pm \infty$, and  also satisfies the optical theorem; as we will
see later, these observations are strongly linked to the success
of the conventional extrapolation-and-regularization procedure for
the vertex function.

Substituting Eq.  (\ref{1-7b}) into Eq.  (\ref{1-6a}) when $\eta
<0$ and into Eqs.  (\ref{1-6a}) and (\ref{1-6b}) when $\eta>0$ and
using $2\lambda=-\eta^2$ one obtains a simple expression for the
vacuo vertex function as
\begin{align}
\label{1-8} \Gamma_0 (E)= \frac{4\pi}{\eta + i \sqrt{E}},
\end{align}
where the branch cut of the square root lies on the real semi-axis
$E>0$, such that on the first Riemann sheet $i\sqrt{-1}=-1$.
Recalling the definition of $E$ one sees that the vacuo vertex
function only depends  on the total momentum and energy of the
molecular pair. Equation (\ref{1-8}) is applicable on both sides of
the resonance and does not have any singularity at the resonance.
On the repulsive side $\eta>0$, Eq. (\ref{1-8}) has a pole at $E = -
\eta^2$, which describes a bound state with binding energy $\eta^2$.
On the attractive side $\eta <0$, the pole is absent: it is located
on the unphysical second Riemann sheet of complex energy $E$.  At
resonance, Eq. (\ref{1-8}) yields a square root singularity, which
is different from the simple pole obtained in Ref.\
\onlinecite{CombescotPRL2003}.

For comparison it is instructive to recalculate $\Gamma_0$ for a
repulsive potential $\eta>0$ using the original Galitskii result
(\ref{1-6a}), i.e., without the bound state. In this case one
obtains a different expression
\begin{align}
\label{1-8b} \Gamma_{0} \left({\bf p}^{\prime}, E\right)  = 4\pi
\left( \frac{2 \eta}{\eta^2 +  {\bf p}^{\prime 2}} - \frac{1}{\eta -
i\sqrt{E}} \right).
\end{align}
Contrary to Eq. (\ref{1-8}), Eq. (\ref{1-8b}) does not have a pole
that corresponds to the bound state. Equations (\ref{1-8}) and
(\ref{1-8b}) explicitly distinguish between potentials that have
the same scattering length but do or do not permit a bound state,
respectively. We note that the series expansion of Eqs.
(\ref{1-8}) and (\ref{1-8b}) with respect to large $\eta$ differs
only in the third-order term. Therefore, both Eqs. (\ref{1-8}) and
(\ref{1-8b}) yield the same result when the Bethe-Salpeter
equation (\ref{1-3}) is solved in a second-order approximation, as
done in the original Galitskii formalism. \cite{FetterWalecka}

We now proceed with the solution of Eq.  (\ref{1-3}) for the
two-particle scattering amplitude (\ref{1-8}).  In doing so, we
perform the analytical continuation defined by Eq.
(\ref{eq:edefinition}). The solution can then be written as
\begin{align}
\label{1-9} \Gamma(g)= \left\{\Gamma_{0}^{-1} + K(g)\right\}^{-1}=
\frac{4\pi}{\eta + i \sqrt{E} + 4\pi K(g)},
\end{align}
where
\begin{align}
& K(g)= \int \frac{d^3{\bf q}}{(2 \pi)^3} \frac{\tanh(\xi_{+}/2T) +
\tanh (\xi_{-}/2T) -2}{2({\bf q}^2-E)}. \label{1-10}
\end{align}
In the limit  $\eta \to \infty$, $\mu\rightarrow -\infty$, where
the system is dominated by the molecular states, the vertex
(\ref{1-9}) reduces to the vacuo vertex function (\ref{1-8}).

In the final step of the derivation we integrate the last term in
Eq. (\ref{1-10}) by parts. This cancels the term $i\sqrt{E}$ in
Eq. (\ref{1-9}), yielding
\begin{subequations}\label{eq:result}
\begin{align}
\label{1-11} &\Gamma(g)= \frac{4\pi}{\eta +{\tilde K}(g)},
\end{align}
where
\begin{align}
&{\tilde K}(g) =  \int \frac{d^3{\bf q}}{4 \pi^2} \left\{ \frac{
\tanh(\xi_{+}/2T) + \tanh(\xi_{-}/2T)}{ {\bf q}^2-E} -
\frac{2}{{\bf q}^2} \right\}. \label{1-12}
\end{align}
\end{subequations}
This expression coincides precisely with the widely used
extrapolated weak-coupling result, \cite{Review1,Review3}
including the ultraviolet regularization for the contact
interaction (see, e.g., Ref.\ \onlinecite{Popov}). The derivation
procedure above demonstrates that in the ladder approximation,
this expression remains strictly valid in the vicinity of the
resonance, as well as on the molecular side $\eta>0$.  The reason
for the fact that the vertex function (\ref{eq:result}),
originally derived in the limit $\eta \to - \infty$, correctly
describes a molecular bound state for $\eta>0$ lies in the fact
that $\Gamma$ approaches $\Gamma_0$ in Eq. (\ref{1-8}). There is
also another, more intuitive reason why the formal extrapolation
of the results for the weakly interacting Fermi limit to the
domain $\eta
>0$ correctly describes the case with a molecular
bound state and not the weakly repulsive Fermi gas. For a purely
repulsive potential without such a state, the resonant limit $\eta
\rightarrow 0$ can only be realized when the effective radius of
the potential also becomes infinite.  This would be in conflict
with the continuity of the vertex function (\ref{eq:result}) as a
function of $\eta$, which also holds across the resonance at
$\eta=0$.

\section{\label{sec:condensate}Vertex function in the condensate state}

The modified Galitskii approach described above can be equally
applied to solve the Bethe-Salpeter equation for the vertex function
in the condensate system. The condensate system has normal as well
as anomalous vertex functions (with four incoming or outgoing lines)
and the resulting Bethe-Salpeter equation becomes a system of
equations \cite{CombescotPRA74}
\begin{align}
&\Gamma(p_1,p_2;p_3,p_4) = u({\bf p}_1-{\bf p}_2) - T \sum_{q_0}
\int \frac{d^{3}{\bf q}}{(2\pi)^{3}} u({\bf q}) \notag \\
& \Big\{ G(p_{1}-q) G(p_{2}+q) \Gamma(p_1-q, p_2+q;p_3,p_4)
\notag \\
&+ F(p_{1}-q)F(p_{2}+q) \Theta (p_1-q, p_2+q;p_3,p_4) \Big\}, \label{2-0} \\
& \Theta(p_1,p_2;p_3,p_4)  = - T \sum_{q_{0}} \int \frac{d^{3}{\bf
q}}{(2\pi)^{3}} u({\bf q}) \notag \\
& \Big\{ G(q-p_1)G(-p_{2}-q) \Theta(p_1-q, p_2+q;p_3,p_4) \notag
\\
&+ F^*(p_{1}-q) F^*(p_{2}-q) \Gamma (p_1-q, p_2+q;p_3,p_4) \Big\},
\label{2-1}
\end{align}
The normal and anomalous Green functions $G$ and
$F$, respectively, are defined as
\begin{align}
&G(q) = \frac{q_{0} + \xi_{\bf q}}{q_{0}^2 - \Delta_{\bf q}^2},~
F(q) = \frac{\Delta}{q_{0}^2 - \Delta_{\bf q}^2}, \label{2-2}
\end{align}
where $\Delta$ is the BCS single particle spectral gap, related to
the condensate density, and
\begin{equation}
\Delta_{\bf q} = \sqrt{\xi_{\bf q}^2 + \Delta^2}, \label{eq:deltap}
\end{equation}
while $\xi_{\bf q}$ is given in Eq. (\ref{1-2}). Solution of this
system along the lines of the Galitskii formalism outlined in the
previous section yields the algebraic system
\begin{align}
& \Gamma(g) = \Gamma_0(g)\{1 -  K(g) \Gamma(g) -
S(g) \Theta(g) \} , \notag \\
& \Theta(g) = - \Gamma_0(-g) \{ K(-g) \Theta(g) +  S^*(g) \Gamma(g)
\},
\label{2-3}
\end{align}
where the kernel $K(g)$ is defined as the integral of Eq.
(\ref{1-4}) over ${\bf k}$ and the Green function, $G$, defined by
Eq. (\ref{2-2}) while $S(g)$ is
\begin{align}
S(g) = \int \frac{d^{3} {\bf q}}{(2\pi)^{3}} T \sum_{k_0}
F\left(\frac{g}{2} - q\right) F\left(\frac{g}{2} + q \right).
    \label{2-3b}
\end{align}
The algebraic pair of Eqs. (\ref{2-3}) is solved by
\begin{align}
\Gamma(g) = \frac{\eta +i\sqrt{E} + 4\pi K(-g)}{\Xi(g)},~ \Theta(g)
= \frac{4\pi S^*(g)}{\Xi(g)},
\label{2-6}
\end{align}
where the common denominator is given by
\begin{align}
\Xi(g) &= (\eta +i\sqrt{E} + 4\pi K(g))(\eta +i\sqrt{E} +
4\pi K(-g)) \notag \\
&-16 \pi^{2} S(g)S^{*}(g). \label{2-7}
\end{align}
As in the normal case the partial integration in the normal kernel
removes $i\sqrt{E}$ and the result for the vertex becomes
equivalent to that in the weakly coupled neutral BCS system, which
again by the virtue of the derivation is valid for arbitrary
$\eta$, including $\eta=0$.

We conclude the analysis by a brief discussion of the gap
equation, which determines the order parameter $\Delta$. According
to the Thouless criterion, the stability of the condensate
requires that the bosonic excitation spectrum defined by the poles
of $\Gamma$ is gapless. This yields
\begin{align}
\eta +i\sqrt{E} + 4\pi K(0) + 4\pi S(0) = 0. \label{2-2-1}
\end{align}
[The second possible equation with the minus sign in front of
$S(0)$ yields instabilities in the corresponding bosonic
excitation spectrum.] The explicit form of Eq. (\ref{2-2-1}) is
\begin{align}
\label{2-2-2}
    -\eta =\frac{2}{\pi} \int_{0}^{\infty} dq
    \left\{\frac{q^{2}}{2\Delta_{q}}\tanh
    \left[ \frac{\Delta_{q}}{2T} \right] - 1
    \right\}.
\end{align}
This is the familiar BCS gap equation for a weakly interacting BCS
system.  However, as the underlying vertex function (\ref{2-6}),
Eq. (\ref{2-2-2}) is valid (within the mean-field approximation)
in the vicinity of the resonance as well as on its repulsive side,
$\eta>0$. The observation that at $\eta>0$ this mean-field gap
equation, combined with the condition of the constant particle
density, correctly describes the bound molecular states (see,
e.g., Refs.\ \onlinecite{meanfield1} and \onlinecite{meanfield4})
is again ensured by the correct limit $\Gamma\to\Gamma_0$ of  the
vertex function in the molecular limit $\eta \to \infty$,
$\mu\rightarrow -\infty$.

\section{\label{sec:conclusions}Conclusions}

In this work we have presented a procedure of solving the
Bethe-Salpeter equation for the two-particle vertex functions in
the ladder approximation in the vicinity of a broad resonance, as
encountered for ultracold Fermi gases close to a Feshbach
resonance. In order to do this we have extended the Galitskii
formalism to account for the molecular states in the interaction
potential and also utilized the universal form of the resonant
scattering amplitude in the vicinity of the resonance. This
allowed for the exact solution of the Bethe-Salpeter equation at
resonance and its vicinity. It is also valid in the
weak-scattering limit, including the limit of the molecular
states. The solution is regular in the BCS-BEC crossover and
coincides exactly with the standard results obtained for the
weakly attractive Fermi gas. This agreement provides a link
between the resonance regime and the weak-interaction regime and
serves to justify the widely used calculational framework in which
the weak-interaction results (with ultraviolet regularization) are
extrapolated to the BCS-BEC crossover and further to the domain of
molecular states.

The derivation of the vertex function can be carried out both in
the normal as well as in the condensate state. As the latter is
used to extract the equation for the condensate density (the gap
equation), the extrapolation procedure for the BCS gap equation
onto the molecular side of the resonance is also perfectly
justified (within the ladder approximation). Thus our results
establish the robustness of the phenomenological extrapolation
from the BCS limit used in many previous works.

We have not discussed the validity of the ladder approximation
itself, which is obviously questionable when the system is close
to the resonance. The applicability of the ladder approximation is
conventionally established for a small gas parameter
$k_F/|\eta|\ll 1$. This condition apparently fails at resonance,
where the gas parameter diverges. Still, it has been found that
the ladder approximation combined with a self-consistency
procedure delivers reliable qualitative, and to some extent even
quantitative information in the complete crossover regime (see,
e.g., Refs.\ \onlinecite{Review1} and \onlinecite{Review3}). The
presented solution of the Bethe-Salpeter equations in the vicinity
of the resonance reveals that the relevant small parameter in the
resonant system is $\propto k_F|f(k_F)|$, where $|f(k_F)|$ is the
absolute value of the scattering amplitude. This quantity
coincides with the conventional gas parameter in the
weak-interaction limit, and, although not small, is not divergent
at resonance, approaching a constant of order unity instead. While
not providing a small expansion parameter, this observation
provides further insight into the apparent successes of the ladder
approximation in studies of the resonant scattering systems.

\acknowledgments

This work was supported by the European Commission, Marie Curie
Excellence Grant MEXT-CT-2005-023778 (Nanoelectrophotonics).

\end{document}